\documentclass[conference,peer-reviewed]{aesconf}

\graphicspath{{./}{figures/}}

\usepackage[utf8]{inputenc}

\usepackage[protrusion=true, expansion=true, tracking=small]{microtype}

\usepackage[numbers,square]{natbib}

\usepackage{booktabs}
\usepackage{color}
\usepackage{url}

\usepackage{stfloats}  
\usepackage{enumitem}  
\usepackage{amsmath}
\usepackage{amssymb}
\usepackage{graphicx}
\usepackage{adjustbox}

\newcommand\linesubsec[1]{\vspace{0.8mm}\noindent\textbf{#1 --- }}

\title{Neutone SDK:\\An Open Source Framework for Neural Audio Processing}

\author[1]{Christopher Mitcheltree}
\author[1]{Bogdan Teleaga}
\author[1]{Andrew Fyfe}
\author[1]{Naotake Masuda}
\author[1]{Matthias Schäfer}
\author[1]{Alfie Bradic}
\author[1]{Nao Tokui}

\affil[1]{Neutone Inc.}

\correspondence{Christopher Mitcheltree}{christopher@neutone.ai}

\lastnames{Mitcheltree et al.}

\shorttitle{Neutone SDK: An Open Source Framework for Neural Audio Processing}

\begin{document}

\twocolumn[
\maketitle 
\begin{onecolabstract}
Neural audio processing has unlocked novel methods of sound transformation and synthesis, yet integrating deep learning models into digital audio workstations (DAWs) remains challenging due to real-time / neural network inference constraints and the complexities of plugin development. 
In this paper, we introduce the Neutone SDK: an open source framework that streamlines the deployment of PyTorch-based neural audio models for both real-time and offline applications. 
By encapsulating common challenges such as variable buffer sizes, sample rate conversion, delay compensation, and control parameter handling within a unified, model-agnostic interface, our framework enables seamless interoperability between neural models and host plugins while allowing users to work entirely in Python. 
We provide a technical overview of the interfaces needed to accomplish this, as well as the corresponding SDK implementations.
We also demonstrate the SDK's versatility across applications such as audio effect emulation, timbre transfer, and sample generation, as well as its adoption by researchers, educators, companies, and artists alike.
The Neutone SDK is available at \url{github.com/Neutone/neutone_sdk}.

\end{onecolabstract}
]

\section{Introduction}
\vspace{-5pt}


Neural audio processing, a research field where machine learning and neural networks are used to analyze, transform, and generate sound, has seen rapid developments in the last decade alongside the increasing popularity of deep learning and the computational efficiency improvements of computers.
Bridging the gap between researchers and practitioners is especially challenging in this field because neural audio processing brings its own set of problems, such as fixed buffer size and sample rate requirements as well as unique streaming and caching properties of neural architectures (see Section~\ref{sec:overview} for more), which are time-consuming to address and require expert domain knowledge to solve effectively.
These problems are in addition to the usual real-time safety and C++ plugin development required for real-time audio processing.
Therefore, when researchers publish new neural audio processing systems, they are often controlled asynchronously through a command line interface, which is in stark contrast to the feedback-driven workflow introduced by audio plugins and the Virtual Studio Technology (VST) standard.

The Neutone SDK is an open source framework for running real-time and non-realtime audio processing neural networks (which we refer to as models) in external audio software applications such as Digital Audio Workstations (DAWs).
The models exported using the SDK are wrapped in a consolidated interface layer that allows for agnostic interoperability across model deployment hosts. 
The Neutone SDK, as well as its counterpart free VST/AU host plugins Neutone FX and Neutone Gen, aim to bridge the gap between neural audio research and the music creator community.
Although the focus of this paper is on the Neutone SDK, it is important to acknowledge that many of the design decisions of the SDK complement the design decisions made for the two host plugin inference engines that run the models exported using the SDK. 

The Neutone project was conceived in early 2022 by Qosmo\footnote{\url{qosmo.jp/en}}, a Tokyo-based AI creativity and music lab, and is now owned and maintained by Neutone\footnote{\url{neutone.ai}}, a company focused on building next-generation AI tools for artists and musicians.
The open source Neutone SDK and accompanying Neutone FX plugin followed on from the RAVE Audition VST project in 2021 (which was limited to deploying a single specific model architecture) to provide a model-agnostic solution for the deployment and distribution of neural audio technologies in the DAW.

Deploying arbitrary audio processing neural networks in a host-agnostic manner requires various interfaces for control parameters, audio processing, DAW compatibility, and real-time / non-realtime inference.
In this paper, we provide an overview of each of these conceptual requirements and how they are implemented in the Neutone SDK with an emphasis on Python-only development and ease of use.


\section{Background}
\vspace{-5pt}


Some notable projects that exist to help bring neural audio research into the DAW include RT-Neural~\cite{chowdhury2021rtneural}, a lightweight solution with a C++ inference library for hard real-time applications. 
RT-Neural's real-time safety constraints make it highly desirable in many audio processing scenarios, however it is limited to handling a subset of operators and model layer types. 
Machine learning frameworks like PyTorch, JAX, and Tensorflow, are at the cutting edge of deep learning research and provide the functionality required to build modern and complex neural networks. 
RT-Neural and similar libraries are not compatible with many model architectures, and a solution built around the external runtimes of these frameworks is often required instead.

Prior to the introduction of the Neutone SDK, researchers interested in productionizing their real-time neural audio models would need to develop a lot of additional infrastructure on top of the model architecture itself to meet the requirements for real-time audio streaming environments. 
To the best of the authors' knowledge, there were no publicly available solutions that streamlined the process of deploying real-time neural audio models to the DAW before the release of the Neutone SDK. 
However, the Neutone SDK drew inspiration from the earlier project Audacitorch that provides a mechanism for deploying non-realtime machine learning models in the Audacity DAW~\cite{garcia2021audacitorch}. 
Audacitorch implements a wrapper design that its model inference counterpart can then interpret within Audacity as part of an integration. 
The Audacity integration allows models to be selected and inferred from within the DAW software where the model outputs are also rendered, thus mitigating the need to leave the DAW environment. 
The Neutone SDK and its host plugins took inspiration from this improved workflow for DAW users, as working within the DAW is familiar and efficient for many and reduces the need to context-switch between software environments when trying to stay focused on the creative process.

Both Neutone SDK and Audacitorch build on top of PyTorch and its model serialization tool TorchScript. 
Serializing a model using TorchScript renders it to a format that does not rely on Python dependencies and can be used in C++ external software applications.
The ease of use, object-orientated design, and utilities of PyTorch, in addition to its popularity among the neural audio research community, make PyTorch and TorchScript the machine learning framework of choice for the Neutone SDK.

Since the release of the Neutone SDK there have been various other tools to improve the state of running neural networks in real-time audio environments like the DAW. 
Similarly to Neutone FX, Anira~\cite{ackva2024anira} is a C++ inference engine catered for the deployment of real-time models built for major framework runtimes like LibTorch, TensorFlow Lite, and ONNX, and can be embedded inside an application without having to build custom infrastructure to support these inference capabilities. 
'nn\_tilde' \footnote{\url{github.com/acids-ircam/nn_tilde}} by IRCAM is another project that allows the deployment of serialized TorchScript models as a MAX/MSP external. 
Historically, nn\_tilde deployed only RAVE models~\cite{caillon2021rave}, but has since demonstrated compatibility with other real-time model architectures. 
Additionally, HARP~\cite{benetatos2025harp} addresses the problem using a different approach by sending and receiving model inputs and outputs via a server to a Python environment and a Gradio interface.
This is ideal for large models that must be hosted remotely on a GPU server and scenarios where there is a stable network connection.
The machine learning tooling of HARP is in Python which is much more familiar to many neural audio processing researchers and means that models can be run from Python directly.
This greatly reduces the complexity for researchers to share their work with the artist community and is why the Neutone SDK is also written in Python, but with an emphasis on real-time and local models running on CPU.


\begin{figure*}[t]
\begin{center}
\vspace{-20pt}
\includegraphics[width=1.0\textwidth]{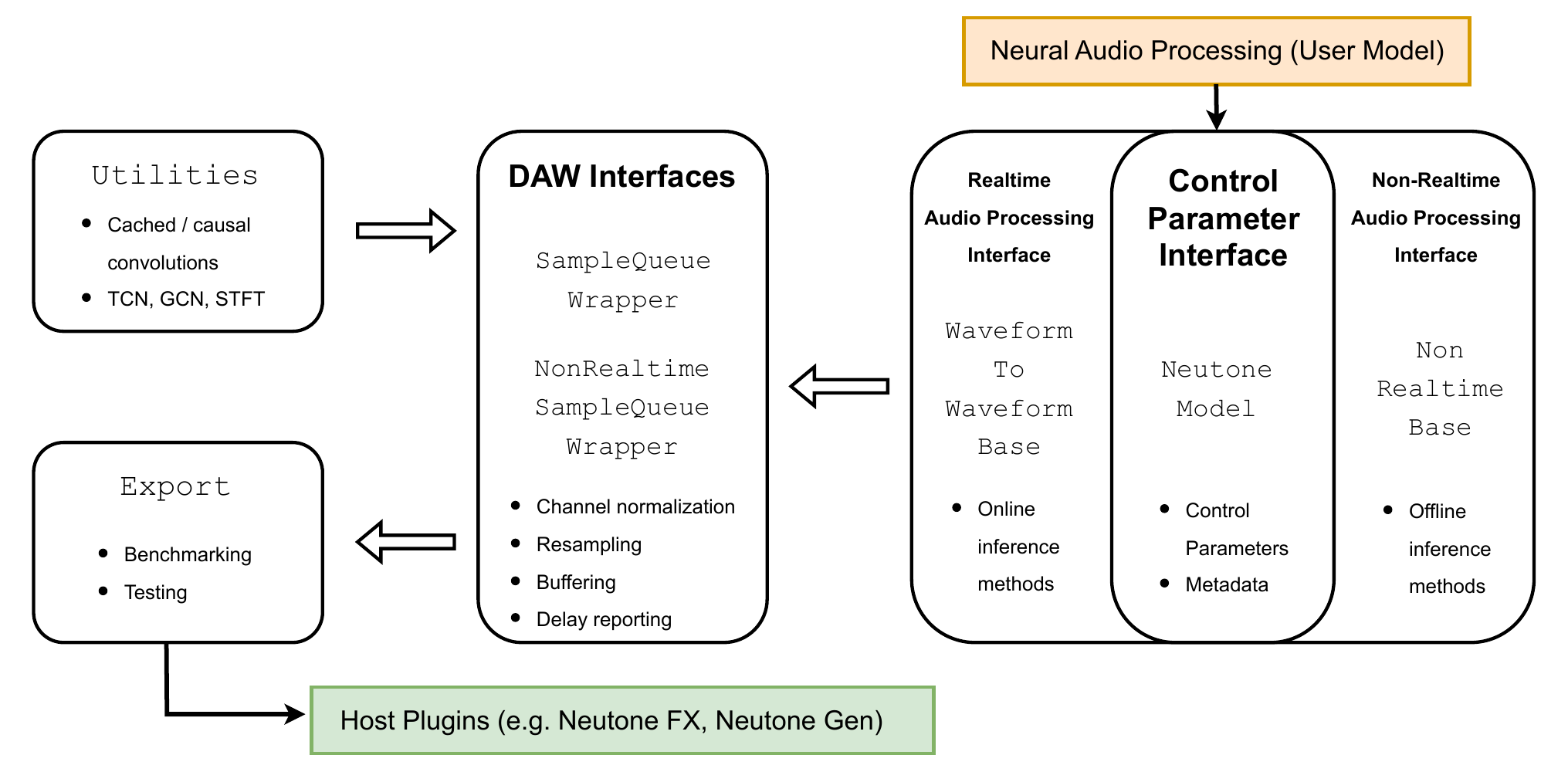}
\vspace{-15pt}
\caption{Overview of the Neutone SDK. Components in \texttt{monospace} font are corresponding SDK implementations.}
\label{fg:sdk_overview}
\end{center}
\vspace{-20pt}
\end{figure*}

\section{Implementation}
\vspace{-5pt}

\subsection{Overview}
\label{sec:overview}
\vspace{-5pt}

The Neutone SDK\footnote{\url{github.com/Neutone/neutone_sdk}} consists of several interfaces and utilities that enable users to wrap, test, benchmark, and export neural audio processing models written in PyTorch so that they can be loaded by a plugin and run in the DAW.
Running neural networks in the DAW can be broken down into several key tasks: handling control parameters, audio processing, dynamic sample rate / channel / buffer size management, and real-time / non-realtime inference.
Each of these tasks is discussed in more detail in the following subsections, and has a corresponding interface in the SDK.
Figure~\ref{fg:sdk_overview} provides an overview of the design of the 
SDK and its components.

All of the classes in the SDK are compatible with TorchScript, meaning they can be exported to C++ automatically, thus enabling the user to work entirely in Python without needing to learn specialized audio plugin development skills.
This also makes it very flexible since users can overwrite and extend any method or class if they have custom requirements for their specific applications. 
In addition to this, the SDK abstracts away many technical challenges that arise when working with real-time audio processing such as:
\begin{itemize}[itemsep=1pt, topsep=1pt]
    \item Variable buffer sizes / optimal buffer size selection
    \item Variable sample rates, optimal sample rate selection, and resampling
    \item Channel normalization
    \item Delay compensation and reporting
    \item Streaming and caching support for real-time audio buffer processing
    \item Control parameter handling and aggregation
    \item Minimizing dynamic memory allocations (since they are not real-time safe)
    \item Lookbehind buffer management
    \item Spectral processing and freq. domain conversion

\end{itemize}
This abstraction is especially important for neural audio processing because many of the aforementioned technical challenges become exacerbated by neural networks, since they typically have fixed sample rate, buffer size, and channel requirements.

\subsection{Neural Audio Processing}
\vspace{-5pt}

Before taking advantage of the Neutone SDK, a user must start with a neural audio processing system that takes audio, control parameters, and/or text as inputs and outputs one or more tracks of audio.
Sections~\ref{sec:examples} and \ref{sec:applications} provide an overview of representative use cases and practical implementations of such systems.

\subsection{Control Parameter Interface}
\vspace{-5pt}

The first interface needed to deploy a neural audio processing system in the DAW is one that defines its control parameters. 
This interface ensures that users can influence model behavior in real time or offline and can be manipulated via external input, such as knobs, sliders, or text fields in a plugin UI.

The Neutone SDK implements this as an abstract base class called \texttt{NeutoneModel} which inherits from \texttt{torch.nn.Module}.
This base class is inherited by \texttt{WaveformToWaveformBase} (Section~\ref{sec:w2w}) and \texttt{NonRealtimeBase} (Section~\ref{sec:nrt}) which are then used to wrap arbitrary \texttt{torch.nn.Module} audio processing neural networks (which we refer to as models).
This is where the names and default values of the control parameters of the model are defined.
The Neutone SDK currently supports three types of control parameters:
\begin{enumerate}[itemsep=1pt, topsep=1pt]
    \item \texttt{ContinuousNeutoneParameter}: a continuous value between 0 and 1 that can be further post-processed by the user and provides control values at a sample or buffer level of granularity. 
    \item \texttt{CategoricalNeutoneParameter}: a categorical value with discrete positions of 2 to $N$ and descriptive labels. 
    \item \texttt{TextNeutoneParameter}: a text input with a maximum number of permitted characters (only available for non-realtime models).
\end{enumerate}
The base class also defines metadata about the model being wrapped that the user must provide, such as its name, authors, tags, description, citation, and technical links which can then be made available to a host plugin.

\subsection{Real-time Audio Processing Interface}
\label{sec:w2w}
\vspace{-5pt}

Next, an interface must be defined for handling input and output audio buffers.
Real-time models generally have a one-to-one mapping between input and output tracks compared to offline models, which is why the Neutone SDK splits these interfaces into corresponding real-time (this section) and non-realtime (Section~\ref{sec:nrt}) implementations.
The \texttt{WaveformToWaveformBase} class (W2W) serves as the real-time audio processing interface in the SDK and inherits from \texttt{NeutoneModel}. 
It wraps an arbitrary \texttt{torch.nn.Module} model that processes sequential audio buffers which means it takes as input an audio tensor of shape ($C_{in}$, $N_{model}$) and outputs an audio tensor of shape ($C_{out}$, $N_{model}$) where $C$ is the number of channels and $N_{model}$ is the number of samples.
Currently, W2W supports values of $C_{in}$ and $C_{out}$ of 1 (mono) and 2 (stereo) as well as any buffer length (i.e., $N_{model} \in \mathbb{Z}^+$).

The \texttt{WaveformToWaveformBase} class contains all audio-specific metadata.
It includes methods for defining $C_{in}$ and $C_{out}$ of the model being wrapped, all supported $N_{model}$ values, what sample rates the model is compatible with ($f_{model}$), and whether / how control parameter values are aggregated from a sample to buffer level of granularity.
If the user's model introduces delay to the input signal, this can be specified and is then made available for the host plugin to query and communicate to the DAW which then compensates the delay to keep all tracks in sync.
Finally, the user can also specify $N_{LB}$ samples of lookbehind buffer which changes the shape of the input audio tensor from 
($C_{in}$, $N_{model}$) to ($C_{in}$, $N_{LB}$ + $N_{model}$).
This is useful for neural networks with a large temporal receptive field that could benefit from looking at previously processed audio samples, but where introducing additional delay to the system via a lookahead buffer is undesirable.

\subsection{Real-time DAW Interface}
\label{sec:sqw}
\vspace{-10pt}

Once a neural model has been equipped with a control parameter interface and an audio processing interface, it must still be adapted to operate within the constraints of a DAW. 
This final layer of abstraction is handled by the real-time DAW interface, which bridges the gap between the wrapped model and the host environment. 
Unlike the previous two interfaces that focus on model behavior and input / output audio, this interface is responsible for managing runtime constraints imposed by the DAW, such as varying buffer sizes, sample rates, and channel configurations. 
The Neutone SDK encapsulates this functionality in the \texttt{SampleQueueWrapper} class which addresses many of the technical challenges listed in Section~\ref{sec:overview}.
It silently handles any mismatches between the DAW and model buffer sizes ($N_{DAW}$ and $N_{model}$), channel dimensions ($C_{DAW}$, $C_{in}$, and $C_{out}$), and sample rates ($f_{DAW}$ and $f_{model}$). 
Normal use of the SDK does not involve interacting with this class, but the SQW is automatically wrapped around W2W classes during the export process (see Section~\ref{sec:export}).

When loading an exported Neutone model, the SQW first determines the optimal buffer size $N_{model}$ and sample rate $f_{model}$ for the wrapped model based on the metadata provided by the user in the audio processing interface and the output audio requirements $N_{DAW}$ and $f_{DAW}$ specified by the DAW and communicated to the SQW by the host plugin.
It also calculates any additional buffering delay caused by mismatched DAW and model buffer sizes.
These calculations are non-trivial and often non-intuitive because they depend on the cyclical properties of the modular arithmetic between the two different buffer sizes.
Resampling between commonly used sample rates such as 48 kHz and 44.1 kHz can also result in coprime resampled input and model buffer sizes which means it is critical to compute the buffering delay correctly to avoid dropouts and stuttering during playback.
We implement the algorithm described in~\cite{letz2021callback} which guarantees the lowest possible amount of additional delay from buffer size adaptation (which we refer to as buffering delay).
Once all buffer sizes and sample rates have been calculated and selected, the SQW pre-allocates input, output, control parameter, and lookbehind circular queues as required and reports the total resampling, buffering, and model delay to the host plugin.

\begin{figure*}[t]
\begin{center}
\vspace{-15pt}
\includegraphics[width=1.0\textwidth]{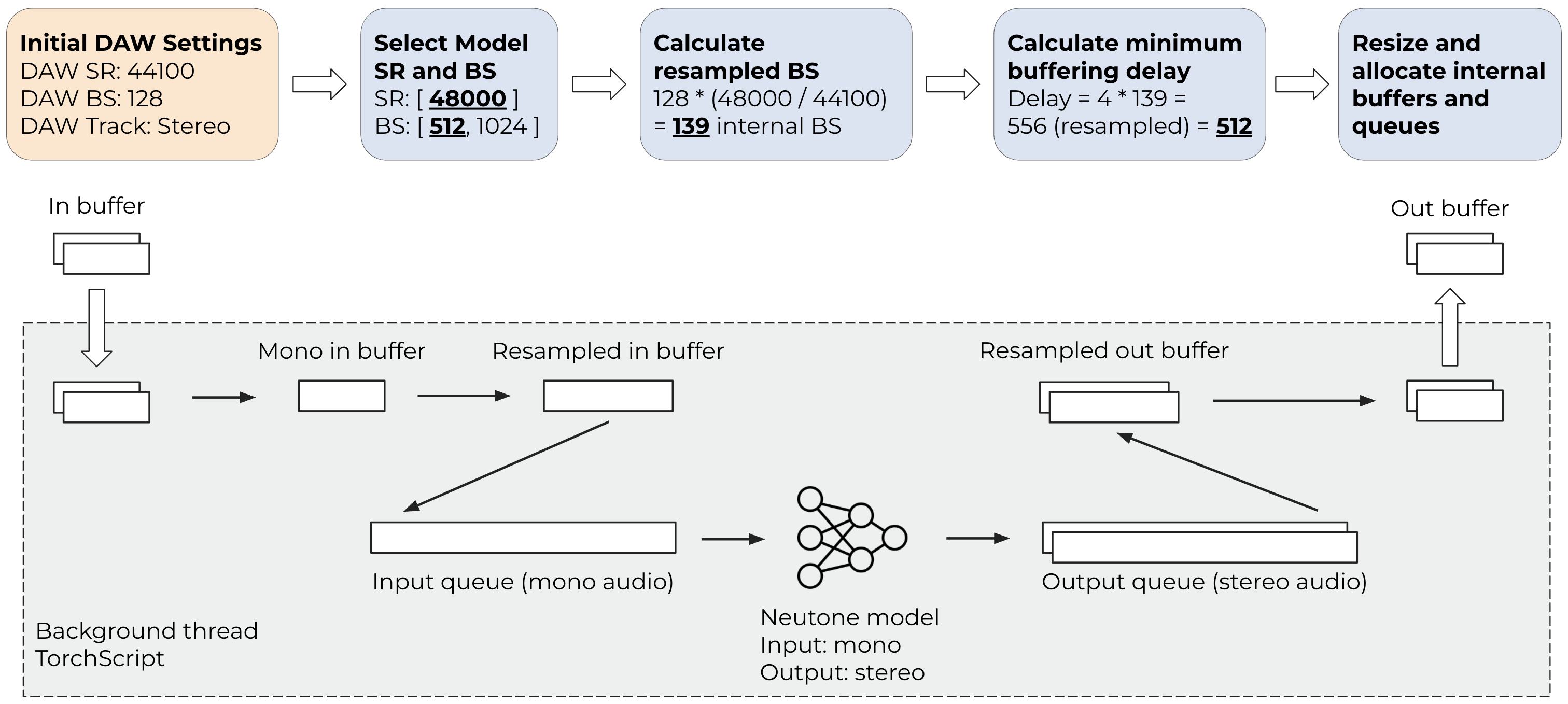}
\vspace{-15pt}
\caption{Flow of an audio buffer through the Neutone SDK. The sequence at the top visualizes the steps that occur when an exported Neutone model is loaded by a host plugin such as Neutone FX. Everything inside the gray box has been compiled to C++ automatically via TorchScript and is handled by the \texttt{SampleQueueWrapper}, \texttt{WaveformToWaveformBase}, and \texttt{NeutoneModel} classes in the SDK.}
\label{fg:buffer_flow}
\end{center}
\vspace{-20pt}
\end{figure*}

During inference, the SQW makes use of the channel normalization and resampling sandwiches (see Section~\ref{sec:sandwiches}) to convert incoming and outgoing audio buffers to the specified model input / output channel sizes ($C_{DAW} \to C_{in}$ and $C_{out} \to C_{DAW}$) and selected optimal model sample rate ($f_{DAW} \to f_{model} \to f_{DAW}$).
It also provides methods for the host plugin to perform audio thread, background thread, and offline inference, each of which require different buffering implementations and overall make the Neutone SDK compatible with most host plugin architectures and audio processing use cases.
Finally, the SQW can adapt the wrapped model to dynamically changing $f_{DAW}$ and $N_{DAW}$ through its \texttt{set\textunderscore daw sample\textunderscore rate\textunderscore and\textunderscore buffer\textunderscore size()} method and it also includes a \texttt{reset()} method for clearing all model, audio processing interface, and real-time DAW interface state.
Figure~\ref{fg:buffer_flow} visualizes the flow of an audio buffer through the SDK, as well as showing an example initialization sequence that occurs when $f_{DAW}$ and $N_{DAW}$ change.

\subsubsection{Neutone FX}
\vspace{-5pt}

\begin{figure}[t]
\begin{center}
\includegraphics[width=1.0\columnwidth]{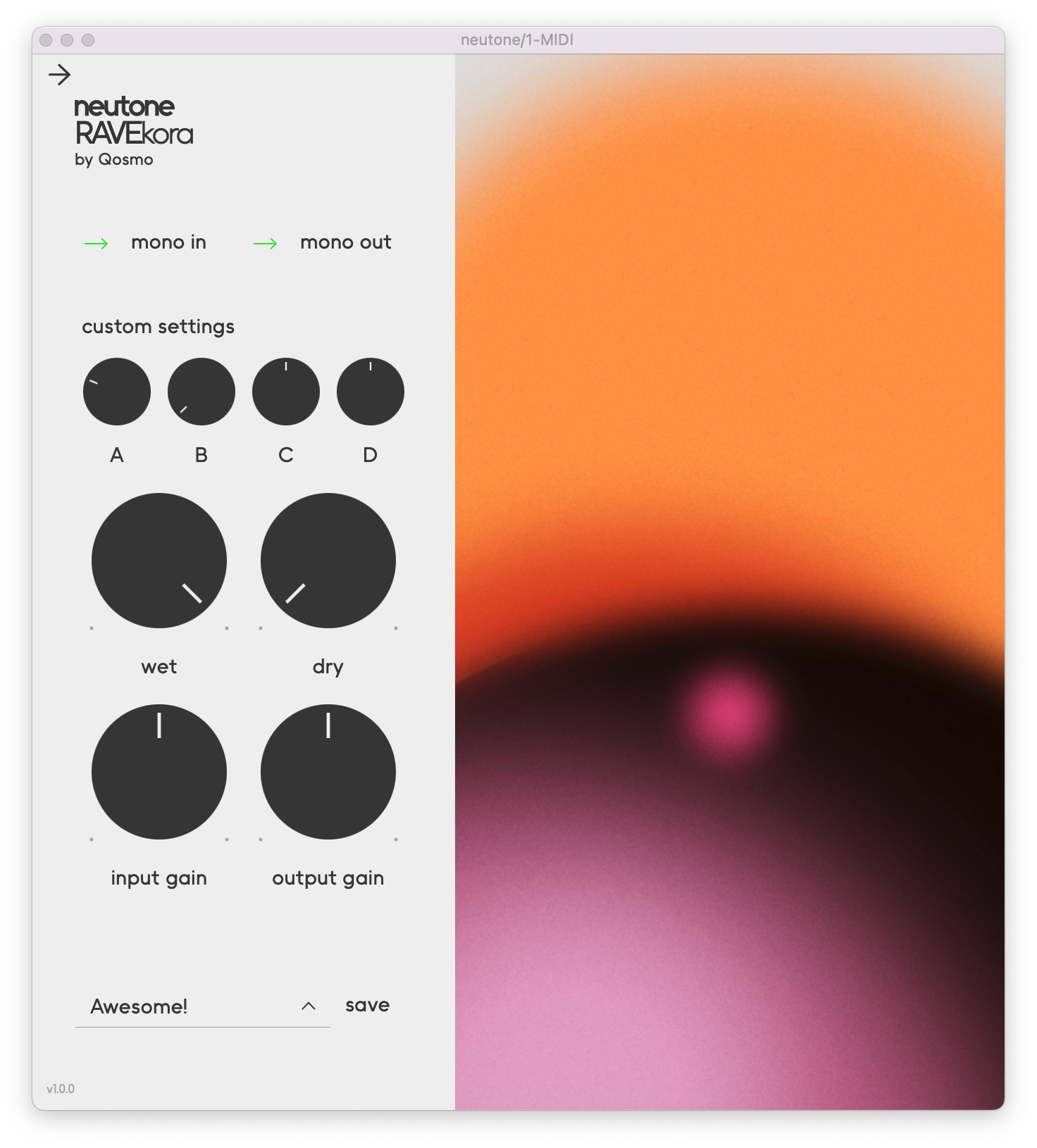}
\vspace{-20pt}
\caption{User interface of the Neutone FX host plugin.}
\label{fg:fx_ui}
\end{center}
\vspace{-20pt}
\end{figure}

Neutone FX\footnote{\url{neutone.ai/fx}}, shown in Figure~\ref{fg:fx_ui}, is a free VST/AU host plugin for MacOS and Windows that was developed alongside the Neutone SDK and provides an interface for artists, creators, and researchers to access real-time neural audio processing models.
It includes a model browser that allows one to filter, search through, and download user models that have been shared and uploaded to the Neutone servers as well as providing the option to load your own locally wrapped and exported model via the SDK (see Section~\ref{sec:export} and Figure~\ref{fg:sdk_workflow}).

Neutone FX is written in C++ using the JUCE framework, and is built as an audio effect, meaning it takes individual audio buffers as input, and processes / outputs them sequentially in a streaming fashion.
Its UI supports four custom control parameter knobs (\texttt{ContinuousNeutoneParameter} and \texttt{CategoricalNeutoneParameter}), as well as wet, dry, input gain, and output gain controls.
The plugin interfaces with the SQW and handles moving audio buffers between the audio (foreground) thread and a background thread, which is required for running neural audio processing models with fixed buffer sizes.
This thread management challenge cannot be solved by the Neutone SDK alone, as the SDK is written in TorchScript compatible PyTorch, limiting direct control over multi-threading. 
This constraint is an intentional design choice to ensure the SDK remains entirely in Python.
Fortunately, open-source frameworks like Anira~\cite{ackva2024anira} provide solutions for efficient real-time processing at the C++ and audio plugin level, making them a natural complement to the Neutone SDK.

\subsection{Non-realtime Audio Processing Interface}
\label{sec:nrt}
\vspace{-5pt}

Offline audio processing models generally have fewer constraints compared to real-time models, but are consequently more versatile and have a wider variety of inputs and outputs.
The \texttt{NonRealtimeBase} class (NRB) in the SDK provides the interface for non-realtime audio processing models.
It inherits from \texttt{NeutoneModel} and wraps an arbitrary \texttt{torch.nn.Module} audio processing model for offline inference.

The NRB is very similar to the W2W class (Section~\ref{sec:w2w}), except that it has been generalized to support multiple input and output audio tracks as well as text control parameters.
This improves the compatibility of the SDK with the comparatively wider range of non-realtime model applications.
A common use case that requires multi-track functionality is stem separation, where a model takes a single input audio track and extracts the individual instrument tracks from it.
The NRB also includes methods for early termination of model inference calls and querying inference progress, which enables updating host plugin UIs accordingly.

\subsection{Non-realtime DAW Interface}
\label{sec:nr_sqw}
\vspace{-5pt}

Just like real-time models, offline models also require an interface to seamlessly handle dynamic host conditions.
The \texttt{NonRealtimeSampleQueueWrapper} (NRSQW) provides this interface for models wrapped with \texttt{NonRealtimeBase}, offering functionality analogous to the real-time \texttt{SampleQueueWrapper}, but optimized for offline processing.
The NRSQW provides the same buffer size, sample rate, and channel normalization as outlined in Section~\ref{sec:sqw}, generalized to multi-track audio inputs and outputs.
During inference, inputs are padded, block-segmented, and processed sequentially, with delay compensation, progress tracking, and cancellation support for responsive host plugin UI updates. 
Outputs are then concatenated and de-padded to preserve alignment with the original input.
The NRSQW supports both one-shot inference as well as block-wise streaming for large inputs, making it well-suited for complex models that exceed real-time constraints.

\subsubsection{Neutone Gen}
\vspace{-5pt}

\begin{figure}[!t]
\begin{center}
\includegraphics[width=1.0\columnwidth]{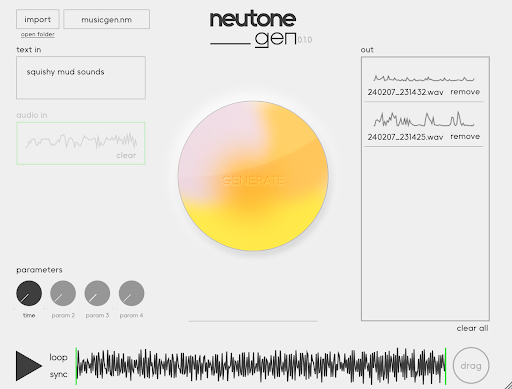}
\vspace{-15pt}
\caption{User interface of the Neutone Gen host plugin.}
\label{fg:gen_ui}
\end{center}
\vspace{-20pt}
\end{figure}

As a natural complement to Neutone FX, Neutone Gen\footnote{\url{neutone.ai/gen}} (shown in Figure~\ref{fg:gen_ui}) is a free VST/AU host plugin for neural audio models that do not run in real time. 
The plugin loosely emulates the design of traditional audio sampler software, whereby users can populate a custom library of audio samples output by the model being hosted. 
These audio samples can be generated, auditioned, and trimmed within the plugin before being dragged into the DAW’s timeline for further processing. 

Neutone Gen’s workflow enables a wider variety of models to be wrapped with the SDK, including those with text prompt input configurations. 
Heavier, high-quality variants of models used in Neutone FX can also be deployed, such as intensive stem separation and timbre transfer algorithms. 
The plugin can interface with both the \texttt{SampleQueueWrapper} and the \texttt{NonRealtimeSampleQueueWrapper} interfaces, thus making all real-time models for Neutone FX compatible with Neutone Gen. 
Audio playback can also be synchronized to the DAW via automatic tempo adjustment and looping controls which facilitates the use of offline models in both music production and live performance scenarios.

\subsection{Additional SDK Functionality}
\vspace{-5pt}

Besides the general interfaces discussed in the previous sections, the SDK also comes with several utility classes and methods to help with building and exporting neural audio processing systems in PyTorch for deployment in the DAW.

\subsubsection{Sandwiches}
\label{sec:sandwiches}
\vspace{-5pt}

The sandwich classes play an integral role in the DAW interface classes and provide implementations of input and output audio channel normalization and resampling algorithms, thus ``sandwiching'' a W2W or NRB wrapped Neutone model in the middle.
Like most components in the Neutone SDK, users can inherit from the existing sandwich classes to implement their own custom algorithms.

\linesubsec{Channel Normalization}
\\the \texttt{ChannelNormalizerSandwich} class is very simple and performs $C_{DAW} \to C_{in}$ and $C_{out} \to C_{DAW}$ mappings by taking the mean of multiple channels when $C_{DAW} > C_{in}$ and $C_{out} > C_{DAW}$, and duplicating channels when $C_{DAW} < C_{in}$ and $C_{out} < C_{DAW}$.



\linesubsec{Resampling}
The \texttt{ResampleSandwich} is an abstract base class defining a bidirectional resampling interface ($f_{\text{DAW}}\to f_{\text{model}}$ and $f_{\text{model}}\to f_{\text{DAW}}$). 
High‑quality, band‑limited resampling is computationally expensive, which presents a challenge for neural audio processing: any cycles spent on filtering and interpolation cannot be spent on expensive neural network inference. 
To address this, the Neutone SDK provides two zero‑allocation, in‑place resamplers:

\begin{enumerate}[itemsep=1pt, topsep=1pt]
    \item \texttt{InplaceLinearResampler}:\\
    a simple linear interpolator (equivalent to convolving with a triangular kernel) that does not apply an anti‑alias filter when downsampling and sacrifices quality for speed and simplicity. 
    This implementation is $\sim$40\% faster than \texttt{torch.nn.functional.interpolate} for common sample‑rate ratios, with zero dynamic memory allocations.
    \item \texttt{Inplace4pHermiteResampler}:\\
    A 4‑point cubic Hermite spline interpolator (informed by~\cite{niemitalo2001polynomial}), also without dynamic memory allocations. 
    The Hermite kernel yields a slightly steeper frequency roll‑off than the triangular kernel, providing mild low‑pass smoothing that reduces (but does not entirely eliminate) aliasing, yet remains only $\sim$1.8\(\times\) slower than the linear resampler. 
    This offers the best quality vs.\ complexity trade‑off among polynomial interpolators on un‑oversampled audio (see~\cite{niemitalo2001polynomial} for more details).
\end{enumerate}


\subsubsection{Utilities}
\vspace{-5pt}

While many neural audio processing researchers are familiar with PyTorch or similar deep learning frameworks, writing TorchScript compatible PyTorch code for real-time inference can be quite challenging, which raises the barrier to entry for using the Neutone SDK.
As a result, the SDK includes several modular, commonly used components for neural audio processing that are optimized, avoid dynamic memory allocations, and are TorchScript-able.

\linesubsec{Circular Queue}
an optimized ring buffer for storing audio or parameter control values and is used extensively by the SQW.

\linesubsec{Cached and Causal Convolution}
cached (streaming) and causal 1D convolution implementations with receptive field reporting for time domain neural audio processing.
Interested readers may refer to~\cite{caillon2022streamable} and~\cite{steinmetz2022efficient} for a detailed overview of this topic.

\linesubsec{TCN}
temporal convolutional neural network (TCN) implementation that supports conditional feature-wise linear modulation (FiLM)~\cite{perez2018film}, streaming, and causal convolutions.
TCNs are a popular architecture for neural audio processing due to their efficiency and large temporal receptive field, making them suitable for real-time applications.
Interested readers may refer to~\cite{steinmetz2022efficient} for an in-depth explanation.

\linesubsec{GCN}
 gated convolutional neural network (GCN)~\cite{comunita2023modelling} implementation: a special case of a TCN for modeling behavior over longer time scales using time-varying FiLM conditioning.

\linesubsec{Spectral Processing}
a streaming implementation (\texttt{RealtimeSTFT} class) of the short-time Fourier transform (STFT) and its inverse for building spectral processing algorithms.
It includes helper methods for handling spectrogram normalization, crossfading, phase information, and delay reporting.
The SDK also includes a \texttt{CachedMelSpec} class which enables streaming computation of Mel spectrogram features.

\linesubsec{Filters}
highpass, lowpass, bandpass, and bandstop streaming finite impulse response (FIR) and infinite impulse response (IIR) filter implementations as well as a state variable filter (SVF) implementation.

\subsubsection{Export}
\label{sec:export}
\vspace{-5pt}

Once a user's model has been wrapped using the audio processing interfaces, models are exported by the SDK as a TorchScript file.
As mentioned in Section~\ref{sec:sqw}, models are automatically wrapped with their corresponding DAW interface (SQW or NRSQW) when exported.
To make it easier for users of the SDK to load models in a host plugin and contribute models upstream, all relevant metadata and input/output example audio samples are bundled inside the TorchScript model file. 
The metadata bundled within the exported model file consists of both functional metadata which is used by the host plugin to perform inference correctly and cosmetic metadata which is used to display information about the model like control knob descriptions and default values in the host plugin UI. 
A schema is used for the metadata to provide consistency for downstream applications.

\begin{figure*}[t]
\begin{center}
\vspace{-20pt}
\includegraphics[width=1.0\textwidth]{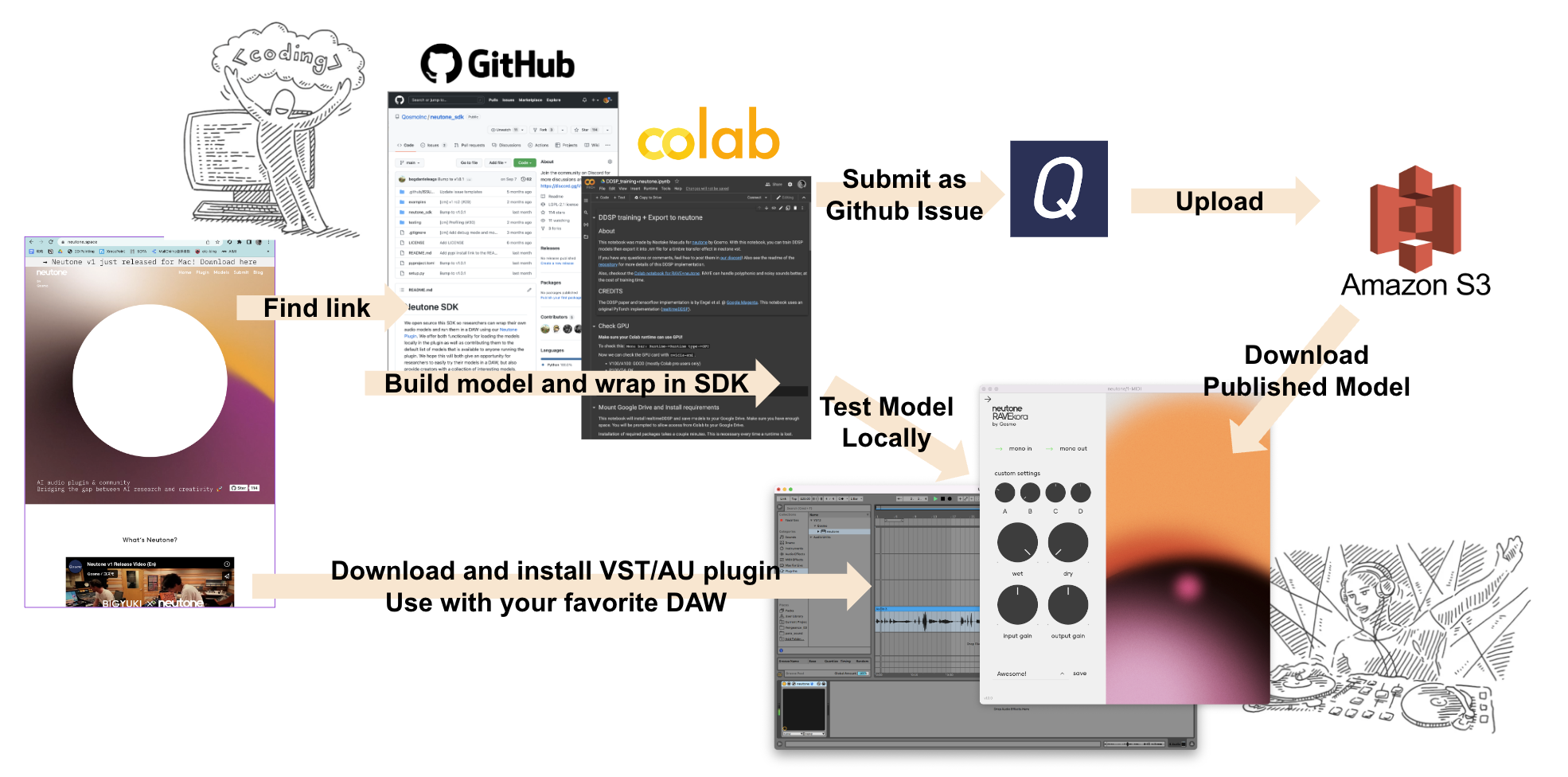}
\vspace{-15pt}
\caption{Workflow for developing, testing, and publishing a neural audio processing model using the Neutone SDK.}
\label{fg:sdk_workflow}
\end{center}
\vspace{-20pt}
\end{figure*}

During export, an additional round of optimization and inference preparation is applied to the model and all Neutone SDK layers before scripting them together.
Once the model is exported to disk, it can be locally loaded into a host plugin like Neutone FX or Gen and tested by the developer in the context of a DAW.
Since the SDK is open source, models can also be loaded and run by third-party host plugins.
The exported file can be shared with anyone, but users can also submit it on GitHub\footnote{\url{github.com/Neutone/neutone_sdk/issues}} to be added to the model browser of Neutone FX, making it available to all users of the plugin around the world.
Models publicized in this way must comply with Neutone's ethics, training data, and copyright policy\footnote{\url{neutone.ai/blog/neutone-on-ai-and-copyright}}.
The entire model development, local testing, export, and publication process is summarized in Figure~\ref{fg:sdk_workflow}.

\linesubsec{Benchmarking}
Two benchmarking utilities are provided within the SDK. 
The speed benchmark can be used to calculate the real-time factor (RTF) of the model on a given machine and verify that the model is fast enough to be executed in real time. 
The latency benchmark can be used to calculate the number of delay samples that will be reported to the DAW for a given model. 
Both benchmarks can be run for different combinations of sample rates and buffer sizes and their results are communicated to the user during export.
A helper method for profiling models is also available and can be used to debug CPU and RAM usage and detect dynamic memory allocations.
    

\subsubsection{Examples}
\label{sec:examples}
\vspace{-5pt}

The Neutone SDK comes with several simple examples and Colab notebooks that can serve as templates for users to get started with the SDK as quickly as possible. 
While the Neutone SDK enables building neural audio processing models without writing any C++ code, the Colab notebook examples take this a step further and allow users to train and export models without writing any code at all.

\section{Applications}
\label{sec:applications}
\vspace{-5pt}

Since the release of the Neutone SDK in early 2022, there has been steady adoption from the community, with the Neutone Discord server\footnote{\url{discord.com/invite/r6WwYCvJTS}} containing more than 2000 members as of writing.  
Users range from independent artists that aim to train models on their own data, to researchers using the SDK for expanding the reach of their published papers, to professors and teachers using it in their classes for educational purposes, to companies using it for rapid internal prototyping before investing in custom C++ plugin development. 
Since any audio-to-audio model can be wrapped with the SDK, the possibilities are endless.
Some common model types are: audio effect emulation / inversion / removal, timbre and tone transfer, stem separation, controllable sample generation, and spectral processing.

Papers that use the Neutone SDK to showcase and make their models accessible in the DAW include: a neural fuzz effect~\cite{comunita2023modelling}, a DDSP-based vocoder~\cite{sudholt2023vocal}, causal modeling of time-varying phaser, flanger, and chorus audio effects~\cite{mitcheltree2023modulation}, a neural spring reverb~\cite{papaleo2024evaluating}, a time-varying differentiable filter evaluated on the Roland TB-303 Bass Line synthesizer~\cite{ycy2024diffapf}, and a DDSP wavetable synthesizer for modulation discovery~\cite{mitcheltree2025modulation}.
Several papers have also been made accessible in the DAW using the Neutone SDK after publication such as MelGAN~\cite{kumar2019melgan}, DDSP~\cite{engel2020ddsp}, RAVE~\cite{caillon2021rave}, randomized overdrive neural networks~\cite{steinmetz2021randomized}, steerable discovery of neural audio effects~\cite{steinmetz2021steerable}, MusicGen~\cite{copet2023musicgen}, NoiseBandNet~\cite{barahona2024noisebandnet}, and more.
One benefit of publicizing academic research models through the Neutone SDK is that others can determine the limits of the system, even if they are not explored in the original academic work. 
In addition to this, allowing users to provide their own inputs reduces the cherry-picking bias of pre-selected audio samples.

Despite all the academic and industry neural audio processing research that leverages the Neutone SDK, the majority of users of the SDK are artists who are incorporating the Neutone host plugins in their creative process and performances.\footnote{\url{neutone.ai/artists}}
At a time when many AI companies are directly competing with artists and training on their work without permission, the Neutone SDK democratizes this nascent technology and serves as a foundation for AI-powered tools that enhance, rather than replace, human creativity and artistic expression.



\section{Conclusion}
\vspace{-6pt}

In this paper, we have presented the interfaces required for real-time and offline neural audio processing in the DAW, a technical overview of the Neutone SDK, and discussed its applications and users.
Development of the SDK is ongoing and we hope to explore the following improvements and directions in the future:

\vspace{-5pt}
\begin{itemize}[itemsep=1pt, topsep=1pt]
    \item New and updated examples and Colab notebooks
    \item Switching from TorchScript to ONNX to improve efficiency and cross-framework compatibility
    \item Implementing more utility components and layers as the neural audio processing field progresses
    \item Supporting symbolic models and releasing the Neutone MIDI SDK prototype
\end{itemize}
\vspace{-5pt}

Readers interested in contributing are encouraged to open a pull request or submit an issue through the Neutone SDK GitHub repository. 

\section*{Acknowledgments}
\vspace{-6pt}

We would like to thank Robin Jungers and Akira Shibata for their significant contributions to Neutone during its formation, as well as Nic Becker and Julian Lenz for their work on the Neutone MIDI SDK prototype.
\vspace{-5pt}

\small
\bibliographystyle{jaes}
\bibliography{refs}
\vspace{-20pt}

\end{document}